\documentclass[fleqn,pra]{revtex4-2}

\usepackage{amsmath,graphicx}
\usepackage{verbatim}
\usepackage{hyperref}

\begin{document}

\title{Variational studies of ro-vibrational spectra of DT$^+$ and T$_2^+$ ions.}

\author{D.T.~Aznabayev$^{1,2,3}$, D.B.~Zhomartova$^{3}$, A.K.~Bekbaev$^{2,3}$, }
\author{Vladimir I. Korobov$^{1}$}
\affiliation{$^1$Bogoliubov Laboratory of Theoretical Physics, Joint Institute for Nuclear Research, Dubna 141980, Russia,}
\affiliation{$^2$The Institute of Nuclear Physics, Ministry of Energy of the Republic of Kazakhstan, 050032 Almaty, Kazakhstan}
\affiliation{$^3$Al-Farabi Kazakh National University, 050038 Almaty, Kazakhstan}

\begin{abstract}
In this work we study the last two hydrogen molecular ion isotopologues: DT$^+$ and T$_2^+$, for which high-precision calculations have not yet been made. We obtain the nonrelativistic solutions of the Schr\"odinger equation for the wide range of vibrational states up to the highest possible vibrational $1s\sigma_g$ state which are of spectroscopic precision. Transition amplitudes for electric dipole transitions, leading-order relativistic corrections and coefficients of the effective hyperfine structure Hamiltonians for both DT$^+$ and T$_2^+$ are also calculated.
\end{abstract}

\maketitle

\section{Introduction}

The tritium-bearing isotopologues HT$^+$, DT$^+$ and T$_2^+$ have been recently considered in context of improving of the fundamental constants such as the Rydberg constant, electron-to-nuclear mass ratio and electric root-mean-square charge radii of the hydrogen isotopes \cite{Karr25}. It is expected that trapping, cooling and spectroscopy of HT$^+$, DT$^+$, and T$_2^+$ can be implemented in the same way as has already been done for HD$^+$ \cite{Alighanbari20,Patra20,Kortunov21} and H$_2^+$ \cite{Alighanbari25}. Of particular interest is the precise determination of the triton charge radius $r_t$ as a stringent test of predictivity in the chiral effective field theory of nuclei \cite{ChEFT}.

The hydrogen molecular ion (HMI) is the simplest stable molecule, which may be studied both theoretically and experimentally with high precision \cite{SchillerCP}. It provides thousands of transitions, which have perfectly narrow spectral lines for Doppler-free \cite{Alighanbari20}, quantum logic spectroscopy \cite{Kienzler24}, nondestructive quantum state control \cite{Heidelberg25}, and for quantum sensing \cite{DeMille24}. Precision studies of molecular hydrogen ions allow stringent tests of quantum electrodynamics and establish new limits on the possible manifestations of new interactions between hadrons, the hypothetical "fifth force" \cite{fifth_force21,Alighanbari23,Delaunay23}.

The article is organized as follows. In Sections II and III we write down the nonrelativistic Schr\"odinger equation for bound-state solutions for the hydrogen molecular ions and the variational expansion which we use in the calculations. For experimental work it is important to know the strength of the various transitions thatcan be induced by laser radiation; the intensities of the spectral lines are discussed in Sec. IV. In Sec. V and VI the leading relativistic corrections to the energy levels of the ro-vibrational states in HMI are discussed. Both spin-independent corrections and hyperfine splitting are considered. The last Section VII presents numerical results, which are the main objective of this paper.

In what follows we assume atomic units: $\hbar=|e|=m_e=1$.

\section{Theory}

Throughout we adhere to the following notation: $v$ is the vibrational quantum number, $L$ is the total orbital angular momentum of the nonrelativistic wave function. The nonrelativistic Hamiltonian in the center of mass frame may be written as:
\begin{equation}\label{H_0}
H_0 
 = \frac{\mathbf{P}_1^2}{2M_1}+\frac{\mathbf{P}_2^2}{2M_2}+\frac{\mathbf{p}_e^2}{2m_e}
      -\frac{Z}{r_1}-\frac{Z}{r_2}+\frac{Z^2}{R},
\end{equation}
where $\mathbf{r}_i = \mathbf{r}_e\!-\!\mathbf{R}_i$ are electron position vectors relative to nuclei and $\mathbf{R} = \mathbf{R}_2\!-\!\mathbf{R}_1$ is the internuclear position vector, $(\mathbf{R}_1,\mathbf{R}_2,\mathbf{r}_e)$ and $(\mathbf{P}_1,\mathbf{P}_2,\mathbf{p}_e)$ are the position vectors and momenta of particles in the center-of-mass frame, $M_1$ and $M_2$ are the masses of the two nuclei, and $Z=1$ is the nuclear charge.

\section{Variational wave function}

The variational bound state wave functions were calculated by solving the three-body Schr\"{o}dinger equation with Coulomb interaction using the
variational approach based on the exponential expansion with randomly chosen exponents \cite{Korobov00}. In this approach, the wave function for a state with a total orbital angular momentum $L$ and of a total spatial parity $\pi=(-1)^L$ is expanded as follows:
\begin{equation}\label{exp_main}
\begin{array}{@{}l}\displaystyle
\Psi_{LM}^\pi(\mathbf{R},\mathbf{r}_1) =
       \sum_{l_1+l_2=L}
         \mathcal{Y}^{l_1l_2}_{LM}({\mathbf{R}},{\mathbf{r}}_1)
         G^{L\pi}_{l_1l_2}(R,r_1,r_2),
\\[4mm]\displaystyle
G_{l_1l_2}^{L\pi}(R,r_1,r_2) =
    \sum_{n=1}^N \Big\{C_n\,\mbox{Re}
          \bigl[e^{-\alpha_n R-\beta_n r_1-\gamma_n r_2}\bigr]
+D_n\,\mbox{Im} \bigl[e^{-\alpha_n R-\beta_n r_1-\gamma_n r_2}\bigr] \Big\},
\end{array}
\end{equation}
here $\mathcal{Y}_{LM}^{l_1 l_2}(\mathbf{x}_1,\mathbf{x}_2)=x_1^{l_1} x_2^{l_2}\{Y_{l_1}(\hat{\mathbf{x}}_1)\otimes Y_{l_2}(\hat{\mathbf{x}}_2)\}_{LM}$ are regular bipolar spherical harmonics \cite{Varsh}, and the complex exponents, $\alpha$, $\beta$, $\gamma$, are generated in a pseudorandom way:
\begin{equation}
\begin{array}{l}\displaystyle
\alpha_i = \left\lfloor\frac{1}{2}i(i+1)\sqrt{p_\alpha}\right\rfloor[(A_2-A_1)+A_1]
   +i\left\lfloor\frac{1}{2}i(i+1)\sqrt{q_\alpha}\right\rfloor[(A'_2-A'_1)+A'_1],
\end{array}
\end{equation}
$\lfloor x\rfloor$ designates the fractional part of $x$,
$p_\alpha$ and $q_\alpha$ are some prime numbers, $[A_1,A_2]$ and
$[A'_1,A'_2]$ are real variational intervals which need to be
optimized. Parameters $\beta_i$ and $\gamma_i$ are obtained in a
similar way.

When "gerade" $1s\sigma_u$ weakly bound and "ungerade" $2p\sigma_u$ vibrational states in T$_2^+$ molecular ion are considered, we use other coordinate geometry and basis expansion:
\begin{equation}
\Psi_{LM}^{\pi P_n}(\mathbf{r}_1,\mathbf{r}_2) =
   \sum_{l_1+l_2=L}
      \mathcal{Y}^{l_1l_2}_{LM}({\mathbf{r}_1},{\mathbf{r}}_2)G^{L\pi}_{l_1l_2}(R,r_1,r_2)
      +P_n\: \mathcal{Y}^{l_1l_2}_{LM}({\mathbf{r}_2},{\mathbf{r}}_1) G^{L\pi}_{l_1l_2}(R,r_2,r_1),
\end{equation}
in order to take into account symmetrization explicitly. Here $P_n$ is the eigenvalue of the permutation operator $\hat{P}_n$ of two identical nuclei: $P_n=(-1)^L$ for "gerade" states, and $P_n=-(-1)^L$ for "ungerade" states. Terms "gerade" (g) and "ungerade" (u) refer to the symmetry of a molecular orbital with respect to inversion of the electron wave function through the center of symmetry.

\section{Transition amplitudes}

In laser experiments it is very important to know the transition intensities. We calculate the reduced matrix elements for spin-averaged transition amplitudes, $d=\langle L'v'\|\mathbf{d}\|Lv \rangle$, of the dipole electric moment operator: $\mathbf{d}=-\mathbf{r}_e+Z(\mathbf{R}_1+\mathbf{R}_2)$,
numerically using the variational wavefunctions (\ref{exp_main}) with basis sets of $N=3000-4000$. Results are presented in the Supplemental Materials (SM) \cite{SM}, and, separately, transitions, which connect the $2p\sigma_u$ "ungerade" and $1s\sigma_g$ "gerade" weakly bound states for the T$_2^+$ ion, are shown in Table \ref{T2plus-dipole}. The data are in atomic units. To obtain the dipole moments in Debye one may use the following expression:
\begin{equation}
\mu_{n'n}(M',M) =
C^{L'M+\mu}_{1\mu;LM}\:\frac{\langle \psi_{n'}\|\mathbf{d}\|\psi_{n}\rangle}{\sqrt{2L+1}}
\times(2.54177 \mbox{ Debye});
\end{equation}
that may be used in evaluation of the intensities of various transitions.

We also calculate the spontaneous emission rate for the transitions. For the dipole transitions this quantity is expressed:
\begin{equation}
A_{nn'}
 = \frac{1}{\tau}\,\frac{4\alpha}{3}\,w_{nn'}^3\,
   \frac{\left|\left\langle\psi_n
      \left\|\mathbf{d}\right\|
   \psi_{n'}\right\rangle\right|^2}{2L+1},
\end{equation}
here $w_{nn'} = E_n-E_{n'}$ and $\tau$ is a unit of time (in atomic units: $\tau=2.41888\times10^{-17}$ s). In all the cases we assume that spontaneous emission proceeds from the upper energy state $n$ to a lower state $n'$.

\section{Spin-independent relativistic corrections}

The leading order relativistic corrections ($R_\infty\alpha^2$) at present are well understood and are described by the Breit-Pauli Hamiltonian \cite{BS}. Here we present in explicit form expressions for different terms, which contribute to this order.

The major contribution comes from the relativistic correction for the bound electron,
\begin{equation}\label{rc}
E_{rc}^{(2)} = \alpha^2\!\left\langle\!
         -\frac{\mathbf{p}_e^4}{8m_e^3}
         +\frac{4\pi Z}{8m_e^2}
           \left[\delta(\mathbf{r}_1)
                +\delta(\mathbf{r}_2)
           \right]
      \right\rangle.
\end{equation}

The other corrections are due to the finite mass of the nuclei and are called recoil corrections of orders $R_\infty\alpha^2(m/M)$, $R_\infty\alpha^2(m/M)^2$, etc. The most important is the transverse photon exchange,
\begin{equation}\label{trans}
\begin{array}{@{}l}
\displaystyle
E_{tr\text{-}ph}^{(2)} =
\frac{\alpha^2Z}{2m_eM_1}
  \left\langle
     \frac{\mathbf{p}_e\mathbf{P}_1}{r_1}
    +\frac{\mathbf{r}_1(\mathbf{r}_1\mathbf{p}_e)\mathbf{P}_1}{r_1^3}
  \right\rangle+
\frac{\alpha^2Z}{2m_eM_2}
  \left\langle
     \frac{\mathbf{p}_e\mathbf{P}_2}{r_2}
    +\frac{\mathbf{r}_2(\mathbf{r}_2\mathbf{p}_e)\mathbf{P}_2}{r_2^3}
  \right\rangle
\\[4mm]\displaystyle\hspace{30mm}
-\frac{\alpha^2Z^2}{2M_1M_2}
  \left\langle
     \frac{\mathbf{P}_1\mathbf{P}_2}{R}
    +\frac{\mathbf{R}(\mathbf{R}\mathbf{P}_1)\mathbf{P}_2}{R^3}
  \right\rangle.
\end{array}
\end{equation}
The contribution of the last term in (\ref{trans}) is not negligible and amounts to about 10\% of $E_{tr\text{-}ph}^{(2)}$.

Tables of numerical expectation values of various operators appearing in Eqs.~(\ref{rc}) and (\ref{trans}) for ro-vibrational states in DT$^+$ and T$_2^+$ ions are presented in SM \cite{SM} (Tables III-V).

\section{Hyperfine structure}

Triton is a particle of spin $1/2$ with the magnetic moment $\boldsymbol{\mu}_t=\mu_t\mu_n\mathbf{I}_t$, while the deuteron is a spin one particle and has the magnetic moment $\boldsymbol{\mu}_d=\mu_d\mu_n\mathbf{I}_d$ and the quadrupole moment $Q_d$. We take dimensionless parameters: $\mu_d=0.857\,438\,23$ and $\mu_t=2.978\,962\,46$, from the CODATA22 recommended values \cite{CODATA22}, and $Q_d=0.285\,699(23)$ fm$^2$ from \cite{Pachucki20}.

The spin dependent part of the Breit-Pauli Hamiltonian for the HMI has the form:
\begin{equation} \label{breit-pauli}
\begin{array}{@{}l}\displaystyle
H_{\rm HFS} = H_{so}+H_{ss}+H_{so_N}+H_{Q_d},
\\[3mm]\displaystyle
H_{so} =
   \frac{Z_a(1\!+\!2a_e)}{2m^2}\,\frac{[\mathbf{r}_a\!\times\!\mathbf{p}_e]}{r_a^3}\,\mathbf{s}_e
   -\frac{Z_a(1\!+\!a_e)}{mM_a}\,\frac{[\mathbf{r}_a\!\times\!\mathbf{P}_a]}{r_a^3}\,\mathbf{s}_e,
\\[3mm]\displaystyle
H_{ss} =
   \left[
      \frac{\boldsymbol{\mu}_e\boldsymbol{\mu}_a}{r_a^3}
      -3\frac{(\boldsymbol{\mu}_e\mathbf{r}_a)(\boldsymbol{\mu}_a\mathbf{r}_a)}{r_a^5}
   \right]
   -\frac{8\pi\alpha}{3}\boldsymbol{\mu}_e\boldsymbol{\mu}_a\delta(\mathbf{r}_a),
\\[3mm]\displaystyle
H_{so_N} =
   \frac{1}{m}\,\frac{[\mathbf{r}_a\!\times\!\mathbf{p}_e]}{r_a^3}\,\boldsymbol{\mu}_a
   -\frac{1}{M_a}\left[1-\frac{Zm_pI_a}{M_a\mu_a}\right] \, \frac{[\mathbf{r}_a\!\times\!\mathbf{P}_a]}{r_a^3}\,\boldsymbol{\mu}_a,
\\[3mm]\displaystyle
H_{Q_d} =
   \frac{Q_d}{2}\left[
      \frac{r_d^2 \mathbf{I}_d^2-3(\mathbf{r}_d\!\cdot\!\mathbf{I}_{d})^2}{r_d^5}
      -\frac{R^2 \mathbf{I}_d^2-3(\mathbf{R}\!\cdot\!\mathbf{I}_{d})^2}{R}
   \right].
\end{array}
\end{equation}
By averaging the spatial variables of the Hamiltonian (\ref{breit-pauli}) one gets the effective Hamiltonian, which depends on the spin and orbital angular momentum variables. It differs in its form for two systems DT$^+$ and T$_2^+$ considered here.

For the case of the DT$^+$ ion it is convenient to choose the effective spin Hamiltonian for hyperfine interaction in the form:
\begin{equation}\label{heff_DT}
\begin{array}{@{}l}
\displaystyle
H_{\rm eff} =
    E_1(\mathbf{L}\!\cdot\!\mathbf{s}_e)
   \!+\!E_2(\mathbf{L}\!\cdot\!\mathbf{I}_t)
   \!+\!E_3(\mathbf{L}\!\cdot\!\mathbf{I}_d)
   \!+\!E_4(\mathbf{I}_t\!\cdot\!\mathbf{s}_e)
   \!+\!E_5(\mathbf{I}_d\!\cdot\!\mathbf{s}_e)
\\[1mm]\displaystyle\hspace{12mm}
   +E_6\Bigl\{2\mathbf{L}^2(\mathbf{I}_t\!\cdot\!\mathbf{s}_e)
          \!-\!3[(\mathbf{L}\!\cdot\!\mathbf{I}_t)(\mathbf{L}\!\cdot\!\mathbf{s}_e)
            \!+\!(\mathbf{L}\!\cdot\!\mathbf{s}_e)(\mathbf{L}\!\cdot\!\mathbf{I}_t)]
       \Bigr\}
\\[1mm]\displaystyle\hspace{12mm}
   +E_7\Bigl\{2\mathbf{L}^2(\mathbf{I}_d\!\cdot\!\mathbf{s}_e)
          \!-\!3[(\mathbf{L}\!\cdot\!\mathbf{I}_d)(\mathbf{L}\!\cdot\!\mathbf{s}_e)
            \!+\!(\mathbf{L}\!\cdot\!\mathbf{s}_e)(\mathbf{L}\!\cdot\!\mathbf{I}_d)]
       \Bigr\}
\\[1mm]\displaystyle\hspace{12mm}
   +E_8\Bigl\{2\mathbf{L}^2(\mathbf{I}_t\!\cdot\!\mathbf{I}_d)
          \!-\!3[(\mathbf{L}\!\cdot\!\mathbf{I}_t)(\mathbf{L}\!\cdot\!\mathbf{I}_d)
            \!+\!(\mathbf{L}\!\cdot\!\mathbf{I}_d)(\mathbf{L}\!\cdot\!\mathbf{I}_t)]
       \Bigr\}
   +E_9\left[\mathbf{L}^2\mathbf{I}_d^2
          \!-\!\frac{3}{2}(\mathbf{L}\!\cdot\!\mathbf{I}_d)
          \!-\!3(\mathbf{L}\!\cdot\!\mathbf{I}_d)^2
       \right].
\end{array}
\end{equation}
The appropriate coupling scheme of the spin and orbital momenta for DT$^+$ ion is
\begin{equation}\label{coupling_DT}
\mathbf{F} = \mathbf{I}_t+\mathbf{s}_e,
\quad
\mathbf{S} = \mathbf{F}+\mathbf{I}_d,
\quad
\mathbf{J} = \mathbf{L}+\mathbf{S}.
\end{equation}

The effective spin Hamiltonian for the T$_2^+$ hyperfine interaction should be taken in the form (as in the case of H$_2^+$ ion):
\begin{equation}\label{effH_T2}
\begin{array}{@{}l}
\displaystyle H_{\rm eff} =
    b_F(\mathbf{I}\cdot\mathbf{s}_e)
   +c_e(\mathbf{L}\cdot\mathbf{s}_e)
   +c_I(\mathbf{L}\cdot\mathbf{I})
\\[2mm]\displaystyle\hspace{12mm}
   +\frac{d_1}{(2L\!-\!1)(2L\!+\!3)}
       \biggl\{
          \frac{2}{3}\mathbf{L}^2(\mathbf{I}\cdot\mathbf{s}_e)
          -[(\mathbf{L}\cdot\mathbf{I})(\mathbf{L}\cdot\mathbf{s}_e)
           \!+\!(\mathbf{L}\cdot\mathbf{s}_e)(\mathbf{L}\cdot\mathbf{I})]
       \biggr\}
\\[0mm]\displaystyle\hspace{12mm}
   +\frac{d_2}{(2L\!-\!1)(2L\!+\!3)}
       \left[
          \frac{1}{3}\mathbf{L}^2\mathbf{I}^2
          -\frac{1}{2}(\mathbf{L}\cdot\mathbf{I})
          -(\mathbf{L}\cdot\mathbf{I})^2
       \right].
\end{array}
\end{equation}
The coupling scheme of angular momentum operators is
\begin{equation}\label{coupling}
\mathbf{F} = \mathbf{I}+\mathbf{s}_e, \qquad \mathbf{J} = \mathbf{L}+\mathbf{F}.
\end{equation}
where $I=I_1+I_2$ is the total nuclear spin, which is equal to 0 when $L$ is even, and to 1 when $L$ is odd. As a consequence, the hyperfine structure is much simpler for the states of even $L$, where only the value $F=1/2$ is allowed.

Transition amplitudes between hyperfine states of a ro-vibrational transition may be expressed via the dipole transition amplitude as follows:
\begin{equation}\label{hfs_transition_intensities}
\left\langle vFSLJ\|\mathbf{d}\|v'F'S'L'J'\right\rangle =
   \delta_{FF'}\delta_{SS'}(-1)^{J+S+L'+1}\sqrt{(2J\!+\!1)(2J'\!+\!1)}
   \left\{\begin{matrix}
      L & 1 & L' \\ J' & S & J
   \end{matrix}\right\}
   \left\langle vL\|\mathbf{d}\|v'L'\right\rangle\,.
\end{equation}
This is a good approximate to the dominant transitions, which conserve the spin configuration of the initial state in the transition. Since the coupled spin states $|FSLJ\rangle$ are not exact eigenvalues (but good approximations) of the effective Hamiltonian $H_{\rm eff}$, after diagonalization weak transitions appear, which connect the states of different spin configurations \cite{Bakalov11}.

\section{Numerical results and discussion}

The main numerical results which are presented here in Tables I-III are the nonrelativistic energies for the ro-vibrational states of DT$^+$ and T$_2^+$ molecular ions. We have managed to get the highest vibrational states with very high precision of twelve significant digits. We use basis sets of size $N=12000-14000$ and octuple precision (about 64 decimal digits) for these calculations. For the case of T$_2^+$ ion there is a permutation symmetry of the identical nuclei that results in appearence of the states which are supported by the ground $1s\sigma_g$ adiabatic molecular potential, and another set of states of "ungerade" symmetry which are supported by the $2s\sigma_u$ adiabatic potential. The "ungerade" states are also presented on Table I.

For the symmetric molecular ions, with two identical nuclei, electric dipole transitions between the states of the same symmetry are fobbiden, but they are allowed for the transitions between "gerade" and "ungerade" states. Intensities of this transitions are given in Table IV. We also show there the nonrelativistic transition frequency. For conversion from the atomic units we use the CODATA22 recommended value \cite{CODATA22} for the Rydberg constant: $R_{\infty}c$=3 289 841 960.25 MHz. We also want to mention here that the spin-spin interaction breaks the $g/u$-symmetry, that is especially important for the weakly bound states near the T$(n=1)$ threshold, that makes possible to induce $E1$ transitions between the states of "gerade" symmetry and also lead to the substantial change in the binding energy of these weakly-bound states. However, this consideration goes beyond the scope of our work and will be considered elsewhere.

Tables \ref{hfsDT-tab} and \ref{hfs-T2tab} provides the hyperfine splitting of the levels for the states of low rotational number $v$ and orbital angular momentum $L$. The coefficients of the effective HFS Hamiltonians (\ref{heff_DT}) and (\ref{effH_T2}) for an extended range of states are given in SM \cite{SM}.

And the last but not least, theoretical and numerical uncertainties. The transition energy is expanded in a series in terms of the fine structure constant $\alpha$: $E(n,n') = E^{(0)}(n,n')+\alpha^2E^{(2)}+\alpha^3E^{(3)}+\dots$. The first term is the nonrelativistic energy, which is the difference of two binding energies: $E^{(0)}(n,n') = E_n-E_{n'}$, and the numerical precision of the calculated data presented in Tables I--III is about $10^{-12}$. The leading order relativistic correction using tabulated values of the operators from SM \cite{SM} may be obtained with the numerical accuracy of $10^{-6}$. The next contribution is the leading order radiative correction and is of magnitude of $\alpha^3\ln{\alpha}\approx10^{-6}$. However, it  may be evaluated with at least two digits accuracy using the formulas from Eqs.~(8), (10), and (11) of Ref.~\cite{Korobov06}, where instead of the Bethe logarithm value $\beta(L,v)$ for a specific ro-vibration state of HMI, the Bethe logarithm value for the ground state of the hydrogen atom is used: $\beta(1S)=2.9841\dots$. Thus, one can obtain the transition energy with a relative \emph{theoretical} error of about $10^{-8}$. Further improvement requires accurate calculation of the nonrelativistic Bethe logarithm for the states of interest, relativistic corrections of order $m\alpha^6$, etc. These calculations go far beyond the scope of our present consideration.

In summary we conclude that a comprehensive set of calculations, which are required for precision spectroscopy of the DT$^+$ and T$_2^+$ hydrogen molecular ions, has been obtained and presented either here in the main text or in the Supplemental Material \cite{SM}. We hope that this theoretical consideration will stimulate new experimental studies, which may bring new valuable information on the properties of the triton.

\section*{Acknowledgements}

This research has been funded by the Committee of Science of the Ministry of Science and Higher Education of the Republic of Kazakhstan (Grant No. BR21881941 and Grant No. AP19175613)

\clearpage

\begin{table}[t]
\caption{Nonrelativistic energies of rovibrational states in the DT$^+$ and $\textmd{T}_2^+$ molecular ion for the total orbital angular momentum $L$=0 and 1 (in a.u).}
\begin{center}
\begin{tabular}{c|@{~~}c@{~~}c@{~~}|@{~~}c@{~~}c@{~~}|c|@{~~}c@{~~}c}
\hline\hline
 & \multicolumn{2}{c}{DT$^+$} & \multicolumn{5}{c}{T$_2^+$} \\
\hline
 & & & \multicolumn{2}{c|}{gerade $1s\sigma_u$ states} & & \multicolumn{2}{c}{ungerade $2p\sigma_u$ states} \\
$v$ & $L=0$  & $L=1$ & $L=0$  & $L=1$ & $~v~$ & $L=0$ & $L=1$  \\
\hline
\vrule width0pt height 12pt
6	& $-$0.563209025756  & $-$0.563118872861  & $-$0.566900094314  & $-$0.566826008263  & 0 & $-$0.499939275591 &$-$0.499937448611  \\
7	& $-$0.558027249031  & $-$0.557940492213  & $-$0.562116379753  & $-$0.562044753840  & 1 & $-$0.499913057123 &$-$0.499912167739  \\
8	& $-$0.553065336032  & $-$0.552981932383  & $-$0.557510711574  & $-$0.557441514328  & 2 & $-$0.499909095063	&                   \\
9	& $-$0.548320604633  & $-$0.548240519158  & $-$0.553080865383  & $-$0.553014069816  &   &                   &                   \\
10	& $-$0.543790876563  & $-$0.543714082374  & $-$0.548824938502  & $-$0.548760522170  &   &                   &                   \\
11	& $-$0.539474483152  & $-$0.539400961734  & $-$0.544741351321  & $-$0.544679296426  &   &                   &                   \\
12	& $-$0.535370274852  & $-$0.535300016428  & $-$0.540828850479  & $-$0.540769143999  &   &                   &                   \\
13	& $-$0.531477634844  & $-$0.531410638842  & $-$0.537086513973  & $-$0.537029147837  &   &                   &                   \\
14	& $-$0.527796497107  & $-$0.527732772758  & $-$0.533513758329  & $-$0.533458729644  &   &                   &                   \\
15	& $-$0.524327369388  & $-$0.524266936479  & $-$0.530110347998  & $-$0.530057659321  &   &                   &                   \\
16	& $-$0.521071361648  & $-$0.521014251454  & $-$0.526876407157  & $-$0.526826066839  &   &                   &                   \\
17	& $-$0.518030220595  & $-$0.517976477065  & $-$0.523812434177  & $-$0.523764456778  &   &                   &                   \\
18	& $-$0.515206371028  & $-$0.515156052264  & $-$0.520919318989  & $-$0.520873725794  &   &                   &                   \\
19	& $-$0.512602964658  & $-$0.512556144799  & $-$0.518198363694  & $-$0.518155183335  &   &                   &                   \\
20	& $-$0.510223936863  & $-$0.510180708502  & $-$0.515651306711  & $-$0.515610575938  &   &                   &                   \\
21	& $-$0.508074071050  & $-$0.508034548371  & $-$0.513280350788  & $-$0.513242115415  &   &                   &                   \\
22	& $-$0.506159068194  & $-$0.506123391217  & $-$0.511088195104  & $-$0.511052511223  &   &                   &                   \\
23	& $-$0.504485613546  & $-$0.504453953694  & $-$0.509078071361  & $-$0.509045006810  &   &                   &                   \\
24	& $-$0.503061417283  & $-$0.503033985278  & $-$0.507253783115  & $-$0.507223419411  &   &                   &                   \\
25	& $-$0.501895161295  & $-$0.501872218099  & $-$0.505619745859  & $-$0.505592180718  &   &                   &                   \\
26	& $-$0.500996137081  & $-$0.500978008456  & $-$0.504181021276  & $-$0.504156371900  &   &                   &                   \\
27	& $-$0.500372786583  & $-$0.500359875352  & $-$0.502943328831  & $-$0.502921736344  &   &                   &                   \\
28	& $-$0.500026961128  & $-$0.500019735172  & $-$0.501912989749  & $-$0.501894625120  &   &                   &                   \\
29	& $-$0.499929521424  & $-$0.499927042612  & $-$0.501096673070  & $-$0.501081744460  &   &                   &                   \\
30	&                    &                    & $-$0.500500506198  & $-$0.500489263735  &   &                   &                   \\
31	&                    &                    & $-$0.500126715608  & $-$0.500119416990  &   &                   &                   \\
32	&                    &                    & $-$0.499958269055  & $-$0.499954785008  &   &                   &                   \\
33	&                    &                    & $-$0.499915889236  & $-$0.499914658168  &   &                   &                   \\
34	&                    &                    & $-$0.499909252544  & $-$0.499909068280  &   &                   &                   \\
\hline\hline
\end{tabular}
\end{center}
\end{table}

\begin{table}[t]
\caption{Nonrelativistic energies of the rovibrational states in  the DT$^+$ molecular ion for $L$=0$-$5 (in a.u).}
\begin{center}
\begin{tabular}{ccccccc}
\hline\hline
\vrule width 0pt height 11pt
$v$    &$L=0$    &$L=1$   &$L=2$     &$L=3$    &$L=4$   &$L=5$   \\
\hline
\vrule width0pt height 12pt
0		&$-$0.599130662836 	&$-$0.599018796391 	&$-$0.598795436108 	  &$-$0.598461323146    &$-$0.598017559212      &$-$0.597465594804   \\
1		&$-$0.592545017185	&$-$0.592436968097	&$-$0.592221232349	  &$-$0.591898530826	&$-$0.591469935039      &$-$0.590936855624   \\
2		&$-$0.586206019863	&$-$0.586101698119	&$-$0.585893407265	  &$-$0.585581848688	&$-$0.585168064864      &$-$0.584653428113   \\
3		&$-$0.580107895403	&$-$0.580007219079	&$-$0.579806209648	  &$-$0.579505549745	&$-$0.579106253937      &$-$0.578609657726   \\
4		&$-$0.574245408532	&$-$0.574148303628	&$-$0.573954427983	  &$-$0.573664446213	&$-$0.573279346063      &$-$0.572800427675   \\
5		&$-$0.568613850020	&$-$0.568520250343	&$-$0.568333376456	  &$-$0.568053875665	&$-$0.567682709967      &$-$0.567221145568   \\

\hline\hline
\end{tabular}
\end{center}

\caption{Nonrelativistic energies of the rovibrational states in  the $\textmd{T}_2^+$ molecular ion for $L$=0$-$5 (in a.u).}
\begin{center}
\begin{tabular}{ccccccc}
\hline\hline
\vrule width0pt height 11pt
$v$    &$L=0$     &$L=1$    &    $L=2$   &$L=3$  & $L=4$ & $L=5$  \\
\hline
\vrule width0pt height 12pt
0		&$-$0.599506910099 	&$-$0.599417152346 	&$-$0.599237876278	     &$-$0.598969558679		&$-$0.598612909551      &$-$0.598168866013  \\
1		&$-$0.593589927776	&$-$0.593502913031	&$-$0.593329117109	     &$-$0.593069005099		&$-$0.592723269554      &$-$0.592292824517  \\
2		&$-$0.587871233599	&$-$0.587786903300	&$-$0.587618470599	     &$-$0.587366389294		&$-$0.587031335086 		&$-$0.586614199739  \\
3		&$-$0.582346606070	&$-$0.582264906322	&$-$0.582101729247	     &$-$0.581857517719		&$-$0.581532931211		&$-$0.581128839970  \\
4		&$-$0.577012172792	&$-$0.576933054273	&$-$0.576775034334	     &$-$0.576538545450		&$-$0.576224231237		&$-$0.575832941244  \\
5		&$-$0.571864401759	&$-$0.571787819652	&$-$0.571634867479	     &$-$0.571405967427		&$-$0.571101748093		&$-$0.570723038975  \\
\hline\hline
\end{tabular}
\end{center}
\end{table}

\begin{table}[t]\label{T2plus-dipole}
\caption{Dipole moments $d = \bigl\langle v'L'\|\mathbf{d}\| vL \bigr\rangle$ (in a.u.), spontaneous transition rates $A_{nn'}$ (in s$^{-1}$) and transition energies (in MHz) of transitions between $1s\sigma_g$ and $2p\sigma_u$ states in $\textmd{T}_2^+$. $a[b]=a\!\times\!10^{b}$}
\begin{center}
\begin{tabular}{c@{\hspace{8mm}}r@{\hspace{10mm}}r@{\hspace{10mm}}r}
\hline\hline
\vrule width0pt height 11pt
$2\pi\sigma_u \rightarrow 1s\sigma_g$ & \multicolumn{3}{c}{$L=0\to L'=1$} \\
\vrule width0pt height 11pt
$v \rightarrow v'$ & dipole moment\hspace*{-4mm} & transition rate\hspace*{-4mm} & transition energy \\
\hline
\vrule width0pt height 12pt
$v(0\rightarrow30)$  & 0.5457  & 1.06128  & 3 618 748.146 \\
$v(0\rightarrow31)$  & 2.3236  & 0.67605  & 1 185 273.453 \\
$v(0\rightarrow32)$  & 6.1831  & 0.00305  &   102 047.054 \\
$v(0\rightarrow33)$  & 2.5259  & 0.00068  &$-$161 974.865 \\
$v(0\rightarrow34)$  & 0.3342  & 2.2[$-$5]&$-$198 754.566 \\[1.5mm]

$v(1\rightarrow30)$  & 0.3011  & 0.37155  & 3 663 297.571 \\
$v(1\rightarrow31)$  & 1.0584  & 0.21084  & 1 285 061.986 \\
$v(1\rightarrow32)$  & 0.2384  & 0.9[$-$4]&   274 556.289 \\
$v(1\rightarrow33)$  &10.5247  & 1.0[$-$5]&    10 534.369 \\
$v(1\rightarrow34)$  & 2.3259  & 2.4[$-$6]& $-$26 245.332 \\[1.5mm]

$v(2\rightarrow30)$  & 0.0649  & 0.01764  & 3 817 326.087 \\
$v(2\rightarrow31)$  & 0.2220  & 0.00983  & 1 383 851.787 \\
$v(2\rightarrow32)$  & 0.0303  & 1.9[$-$6]&   300 625.388 \\
$v(2\rightarrow33)$  & 0.0903  & 3.0[$-$8]&    36 603.468 \\
$v(2\rightarrow34)$  &33.8908  & 0.2[$-$9]&    $-$176.232 \\
\hline\hline
\end{tabular}
\end{center}
\end{table}

\begin{table}[t]
\caption{Dipole moments $d = \bigl\langle v'L'\|\mathbf{d}\| vL \bigr\rangle$ (in a.u.), spontaneous transition rates $A_{nn'}$ (in s$^{-1}$) and transition energies (in MHz) of transitions between $1s\sigma_g$ and $2p\sigma_u$ states in  $\textmd{T}_2^+$.}
\begin{center}
\begin{tabular}{c@{\hspace{8mm}}r@{\hspace{10mm}}r@{\hspace{10mm}}r}
\hline\hline
\vrule width0pt height 11pt
$2\pi\sigma_u \rightarrow 1s\sigma_g$ & \multicolumn{3}{c}{$L=1\to L'=0$} \\
\vrule width0pt height 11pt
$v \rightarrow v'$ & dipole moment\hspace*{-4mm} & transition rate\hspace*{-4mm} & transition energy \\
\hline
\vrule width0pt height 12pt
$v(0\rightarrow30)$  & 0.5207  & 0.34562 &  3 704 740.954 \\
$v(0\rightarrow31)$  & 2.2330  & 0.24138 &  1 245 317.015 \\
$v(0\rightarrow32)$  & 6.1121  & 0.00241 &    136 991.942 \\
$v(0\rightarrow33)$  & 2.8807  & 0.00178 & $-$141 853.870 \\
$v(0\rightarrow34)$  & 0.5480  & 0.00014 & $-$185 521.210 \\[1.5mm]

$v(1\rightarrow30)$  & 0.2844  & 0.11763 & 3 871 081.095 \\
$v(1\rightarrow31)$  & 1.0190  & 0.07321 & 1 411 657.156 \\
$v(1\rightarrow32)$  & 0.4408  & 0.00014 &   303 332.082 \\
$v(1\rightarrow33)$  &10.0818  & 0.00004 &    24 486.271 \\
$v(1\rightarrow34)$  & 3.9262  & 0.00001 & $-$19 182.869 \\[1.5mm]

\hline\hline
\end{tabular}
\end{center}
\end{table}

\begin{table}[t]
\caption{Hyperfine splitting (in MHz) of ro-vibrational levels of the
$\mbox{DT}^+$ molecular ion.} \label{hfsDT-tab}
\begin{tabular}{ccc@{\hspace{1mm}}c@{\hspace{1mm}}c@{\hspace{3mm}}c@{\hspace{3mm}}c@{\hspace{2mm}}c@{\hspace{2mm}}c@{\hspace{3mm}}c@{\hspace{2mm}}c@{\hspace{2mm}}c@{\hspace{2mm}}c@{\hspace{2mm}}c}
\hline\hline
\vrule width0pt height11pt
 & & \multicolumn{3}{c}{$(F,S)=(0,1)$} & $(1,0)$ & \multicolumn{3}{c}{$(1,1)$} & \multicolumn{5}{c}{$(1,2)$} \\
$L$ & $v$ & $J\!=\!L\!+\!1$ & $L$ & $L\!-\!1$ & $L$ & $L\!+\!1$ & $L$ & $L\!-\!1$
& $L\!+\!2$ & $L\!+\!1$ & $L$ & $L\!-\!1$ & $L\!-\!2$ \\
\hline
\vrule width0pt height10pt
0 & 0 & ~$-$754.625 & & & 105.263 & ~187.506 & & & ~319.219 & & & & \\
0 & 1 & ~$-$741.545 & & & 103.441 & ~184.257 & & & ~313.685 & & & & \\
0 & 2 & ~$-$729.109 & & & 101.709 & ~181.168 & & & ~308.423 & & & & \\
0 & 5 & ~$-$695.403 & & & ~97.016 & ~172.796 & & & ~294.161 & & & & \\
\vrule width0pt height10.5pt
1 & 0 & ~$-$755.529 & $-$753.851 & $-$750.786 & 100.618 & ~194.563 & 169.399 & 191.498 & ~322.049 & 335.704 & 294.250 & & \\
1 & 1 & ~$-$742.413 & $-$740.793 & $-$737.855 & ~99.138 & ~191.035 & 166.947 & 188.062 & ~316.411 & 329.478 & 289.514 & & \\
1 & 2 & ~$-$729.941 & $-$728.380 & $-$725.563 & ~97.728 & ~187.675 & 164.629 & 184.794 & ~311.047 & 323.548 & 285.032 & & \\
1 & 5 & ~$-$696.133 & $-$694.737 & $-$692.265 & ~93.898 & ~178.537 & 158.441 & 175.921 & ~296.493 & 307.402 & 273.019 & & \\
\vrule width0pt height10.5pt
2 & 0 & ~$-$756.264 & $-$753.360 & $-$749.722 & ~98.276 & ~200.991 & 176.446 & 171.704 & ~328.101 & 336.019 & 316.505 & 288.132 & 262.193 \\
2 & 1 & ~$-$743.115 & $-$740.318 & $-$736.824 & ~96.914 & ~197.201 & 173.554 & 169.241 & ~322.224 & 329.785 & 310.973 & 283.679 & 258.977 \\
2 & 2 & ~$-$730.612 & $-$727.920 & $-$724.565 & ~95.621 & ~193.588 & 170.814 & 166.907 & ~316.630 & 323.848 & 305.717 & 279.472 & 255.959 \\
\hline\hline
\end{tabular}
\end{table}

\begin{table}
\caption{Hyperfine splitting (in MHz) of the ro-vibrational levels in the $\mbox{T}_2^+$ molecular ion. The states are $(F,J)$.}\label{hfs-T2tab}
\vspace{-2mm}
\begin{center}
\begin{tabular}{cc@{\hspace{1mm}}@{\hspace{2mm}}c@{\hspace{2mm}}c@{\hspace{5mm}}c@{\hspace{2mm}}c@{\hspace{2mm}}c@{\hspace{2mm}}       c@{\hspace{2mm}}|@{\hspace{1mm}}cc@{\hspace{1mm}}@{\hspace{2mm}}c@{\hspace{2mm}}c}
\hline\hline \vrule width0pt height11.5pt depth4.5pt $L$ & $v$
 & $\left(\frac{1}{2},L\!+\!\frac{1}{2}\right)$
 & $\left(\frac{1}{2},L\!-\!\frac{1}{2}\right)$
 & $\left(\frac{3}{2},L\!+\!\frac{3}{2}\right)$
 & $\left(\frac{3}{2},L\!+\!\frac{1}{2}\right)$
 & $\left(\frac{3}{2},L\!-\!\frac{1}{2}\right)$
 & $\left(\frac{3}{2},L\!-\!\frac{3}{2}\right)$
 & $L$ & $v$
 & $\left(\frac{1}{2},L\!+\!\frac{1}{2}\right)$
 & $\left(\frac{1}{2},L\!-\!\frac{1}{2}\right)$ \\
\hline \vrule width0pt height10.5pt
1 & 0 & $-$994.4271 & $-$988.3034 & 493.9196 & 528.3114 & 438.7758 & & 2 & 0 & 14.3690 & $-$21.5535 \\
1 & 1 & $-$978.8895 & $-$972.9592 & 486.2798 & 519.3847 & 433.1295 & & 2 & 1 & 13.8551 & $-$20.7827 \\
1 & 2 & $-$964.0356 & $-$958.2943 & 478.9776 & 510.8348 & 427.7633 & & 2 & 2 & 13.3558 & $-$20.0336 \\
\vrule width 0pt height 11pt
3 & 0 & $-$997.3209 & $-$981.6585 & 501.0698 & 525.8978 & 488.7862 & 429.4800 & 4 & 0 & 28.5375 & $-$35.6719 \\
3 & 1 & $-$981.6817 & $-$966.5543 & 493.1826 & 517.0646 & 481.2841 & 424.1829 & 4 & 1 & 27.5158 & $-$34.3948 \\
3 & 2 & $-$966.7290 & $-$952.1224 & 485.6398 & 508.6050 & 474.1176 & 419.1557 & 4 & 2 & 26.5229 & $-$33.1537 \\
\hline\hline
\end{tabular}
\end{center}
\end{table}

\clearpage

\setcounter{table}{0}
\setcounter{section}{0}
\setcounter{equation}{0}

\begin{center}\large\bf
Variational studies of ro-vibrational spectra of DT$^+$ and T$_2^+$ ions.\\Supplemental material.
\end{center}

\section{Transition intensities}

In Tables I and II we present dipole moments and emission rates (in s$^{-1}$). For transitions indicated by arrow directions as going from a lower energy state to an upper state, the emission rate is indicated for the backward transition from the upper state to the lower one and then arrows in the table should be reverted.

\section{Spin-independent relativistic corrections}

Results of numerical calculation of the mean values of various operators encountered in formulas of the manuscript related to the spin-idependent relativistic corrections are presented in Tables \ref{T2plus-av} and \ref{DTplus-av}. We use the following notation for the transverse photon exchange contributions:
\begin{equation}
\begin{array}{@{}l}
\displaystyle
R_{ne} =
  -\left\langle
     \frac{\mathbf{p}_e\mathbf{P}_n}{r_n}
    +\frac{\mathbf{r}_n(\mathbf{r}_n\mathbf{p}_e)\mathbf{P}_n}{r_n^3}
  \right\rangle,
\qquad
R_{nn} =
  -\left\langle
     \frac{\mathbf{P}_1\mathbf{P}_2}{R}
    +\frac{\mathbf{R}(\mathbf{R}\mathbf{P}_1)\mathbf{P}_2}{R^3}
  \right\rangle.
\end{array}
\end{equation}

We also calculate the so-called $Q$ term, which is required for the $R_\infty\alpha^3(m/M)$ order recoil corrections, we denote: $Q_{ne} = Q(r_{ne})$, where the divergent distribution $1/r^3$ is defined as follows:
\begin{equation}\label{Qterm}
Q(r) = \lim_{\rho \to 0} \left\langle
            \frac{\Theta(r - \rho)}{ 4\pi r^3 }
      + (\ln \rho + \gamma_E)\delta(\mathbf{r}) \right\rangle,
\end{equation}
to have a finite expectation value. Here $\gamma_E=0.577216\dots$ is Euler's constant.

\begin{table}
\caption{Dipole moments $d = \bigl\langle v'L'\|\mathbf{d}\| vL \bigr\rangle$ of transitions between states in  DT$^+$  (in a.u.).}
\begin{center}
\begin{tabular}{c@{~~}c@{~~}c@{~~}c@{~~}c@{~~}c}
\hline\hline
\vrule width0pt height 11pt
$v \rightarrow v'$   & $L=(0 \rightarrow 1)$  & $L=(1 \rightarrow 2)$   & $L=(2 \rightarrow 3)$  & $L=(3 \rightarrow 4)$  & $L=(4 \rightarrow 5)$  \\
\hline
\vrule width0pt height 12pt
$v(0 \rightarrow 0)$   & 0.2034035  & 0.2878892  & 0.3530662  & 0.4084533  & 0.4577658	\\
$v(0 \rightarrow 1)$   & 0.0176838  & 0.0241409  & 0.0285150  & 0.0317259  & 0.0341452	\\
$v(0 \rightarrow 2)$   & 0.0019551  & 0.0027389  & 0.0033194  & 0.0037887  & 0.0041824  \\
$v(0 \rightarrow 3)$   & 0.0003801  & 0.0005385  & 0.0006599  & 0.0007617  & 0.0008502  \\
$v(0 \rightarrow 4)$   & 0.0000997  & 0.0001422  & 0.0001755  & 0.0002039  & 0.0002291	\\
$v(0 \rightarrow 5)$   & 0.0000318  & 0.0000456  & 0.0000565  & 0.0000660  & 0.0000745  \\[1.5mm]
$v(1 \rightarrow 0)$   & 0.0189286  & 0.0276604  & 0.0349764  & 0.0416652  & 0.0480201  \\
$v(1 \rightarrow 1)$   & 0.2120035  & 0.3000495  & 0.3679559  & 0.4256405  & 0.4769736  \\
$v(1 \rightarrow 2)$   & 0.0251741  & 0.0343397  & 0.0405287  & 0.0450542  & 0.0484465  \\
$v(1 \rightarrow 3)$   & 0.0033818  & 0.0047344  & 0.0057337  & 0.0065396  & 0.0072139  \\
$v(1 \rightarrow 4)$   & 0.0007517  & 0.0010644  & 0.0013037  & 0.0015037  & 0.0016773  \\
$v(1 \rightarrow 5)$   & 0.0002183  & 0.0003112  & 0.0003838  & 0.0004456  & 0.0005004  \\[1.5mm]
$v(2 \rightarrow 0)$   & 0.0019859  & 0.0028262  & 0.0034798  & 0.0040356  & 0.0045275  \\
$v(2 \rightarrow 1)$   & 0.0269850  & 0.0394597  & 0.0499286  & 0.0595130  & 0.0686301  \\
$v(2 \rightarrow 2)$   & 0.2208147  & 0.3125088  & 0.3832121  & 0.4432518  & 0.4966567  \\
$v(2 \rightarrow 3)$   & 0.0310432  & 0.0423124  & 0.0498973  & 0.0554208  & 0.0595394  \\
$v(2 \rightarrow 4)$   & 0.0047809  & 0.0066883  & 0.0080943  & 0.0092252  & 0.0101689  \\
$v(2 \rightarrow 5)$   & 0.0011773  & 0.0016658  & 0.0020389  & 0.0023502  & 0.0026199  \\[1.5mm]
$v(3 \rightarrow 0)$   & 0.0003774  & 0.0005309  & 0.0006462  & 0.0007409  & 0.0008217  \\
$v(3 \rightarrow 1)$   & 0.0034400  & 0.0048989  & 0.0060360  & 0.0070049  & 0.0078641  \\
$v(3 \rightarrow 2)$   & 0.0333249  & 0.0487635  & 0.0617407  & 0.0736380  & 0.0849689  \\
$v(3 \rightarrow 3)$   & 0.2298569  & 0.3252951  & 0.3988692  & 0.4613268  & 0.5168596  \\
$v(3 \rightarrow 4)$   & 0.0361000  & 0.0491657  & 0.0579303  & 0.0642861  & 0.0689992  \\
$v(3 \rightarrow 5)$   & 0.0061762  & 0.0086339  & 0.0104412  & 0.0118913  & 0.0130979  \\[1.5mm]
$v(4 \rightarrow 0)$   & 0.0000976  & 0.0001364  & 0.0001649  & 0.0001877  & 0.0002067  \\
$v(4 \rightarrow 1)$   & 0.0007474  & 0.0010523  & 0.0012818  & 0.0014706  & 0.0016322  \\
$v(4 \rightarrow 2)$   & 0.0048701  & 0.0069404  & 0.0085573  & 0.0099378  & 0.0111644  \\
$v(4 \rightarrow 3)$   & 0.0388109  & 0.0568302  & 0.0720014  & 0.0859298  & 0.0992112  \\
$v(4 \rightarrow 4)$   & 0.2391525  & 0.3384399  & 0.4149663  & 0.4799110  & 0.5376333  \\
$v(4 \rightarrow 5)$   & 0.0406575  & 0.0553273  & 0.0651343  & 0.0722149  & 0.0774349  \\[1.5mm]
$v(5 \rightarrow 0)$   & 0.0000308  & 0.0000429  & 0.0000516  & 0.0000585  & 0.0000641  \\
$v(5 \rightarrow 1)$   & 0.0002141  & 0.0002993  & 0.0003620  & 0.0004125  & 0.0004546  \\
$v(5 \rightarrow 2)$   & 0.0011722  & 0.0016515  & 0.0020131  & 0.0023113  & 0.0025672  \\
$v(5 \rightarrow 3)$   & 0.0063003  & 0.0089851  & 0.0110861  & 0.0128837  & 0.0144839  \\
$v(5 \rightarrow 4)$   & 0.0437769  & 0.0641467  & 0.0813255  & 0.0971194  & 0.1121985  \\
$v(5 \rightarrow 5)$   & 0.2487272  & 0.3519799  & 0.4315481  & 0.4990564  & 0.5590364  \\
\hline\hline
\end{tabular}
\end{center}
\end{table}

\begin{table}[h]
\caption{Spontaneous transition rates (in s$^{-1}$) for transitions (from upper to lower) between states in  DT$^+$.}
\begin{center}
\begin{tabular}{c@{~~}c@{~~}c@{~~}c@{~~}c@{~~}c}
\hline\hline
\vrule width0pt height 11pt
$v \rightarrow v'$   & $L=(0 \rightarrow 1)$  & $L=(1 \rightarrow 2)$   & $L=(2 \rightarrow 3)$  & $L=(3 \rightarrow 4)$  & $L=(4 \rightarrow 5)$  \\
\hline
\vrule width0pt height 12pt
$v(0 \rightarrow 0)$   & 0.0004135  & 0.0039565  & 0.0142270  & 0.0346990  & 0.0686194 \\
$v(0 \rightarrow 1)$   & 0.6696527  & 0.7841869  & 0.8162663  & 0.8186494  & 0.8059642 \\
$v(0 \rightarrow 2)$   & 0.0603604  & 0.0726702  & 0.0777888  & 0.0802523  & 0.0813008 \\
$v(0 \rightarrow 3)$   & 0.0072132  & 0.0088102  & 0.0095673  & 0.0100130  & 0.0102902 \\
$v(0 \rightarrow 4)$   & 0.0011066  & 0.0013645  & 0.0014959  & 0.0015803  & 0.0016393 \\
$v(0 \rightarrow 5)$   & 0.0002073  & 0.0002574  & 0.0002842  & 0.0003023  & 0.0003157 \\[1.5mm]

$v(1 \rightarrow 0)$   & 2.0822437  & 1.4043412  & 1.2734269  & 1.2170729  & 1.1828648 \\
$v(1 \rightarrow 1)$   & 0.0004048  & 0.0038726  & 0.0139224  & 0.0339474  & 0.0671106 \\
$v(1 \rightarrow 2)$   & 1.2104197  & 1.4154202  & 1.4710632  & 1.4729346  & 1.4475632 \\
$v(1 \rightarrow 3)$   & 0.1609398  & 0.1934955  & 0.2068317  & 0.2130724  & 0.2155341 \\
$v(1 \rightarrow 4)$   & 0.0251228  & 0.0306444  & 0.0332336  & 0.0347351  & 0.0356487 \\
$v(1 \rightarrow 5)$   & 0.0047179  & 0.0058102  & 0.0063614  & 0.0067119  & 0.0069535 \\[1.5mm]

$v(2 \rightarrow 0)$   & 0.1776941  & 0.1166487  & 0.1029783  & 0.0958418  & 0.0907297 \\
$v(2 \rightarrow 1)$   & 3.7733566  & 2.5477742  & 2.3126713  & 2.2124253  & 2.1520920 \\
$v(2 \rightarrow 2)$   & 0.0003953  & 0.0037808  & 0.0135899  & 0.0331283  & 0.0654699 \\
$v(2 \rightarrow 3)$   & 1.6389077  & 1.9137008  & 1.9858239  & 1.9850074  & 1.9472930 \\
$v(2 \rightarrow 4)$   & 0.2861029  & 0.3434984  & 0.3666460  & 0.3771491  & 0.3809238 \\
$v(2 \rightarrow 5)$   & 0.0547464  & 0.0666891  & 0.0722257  & 0.0753859  & 0.0772624 \\[1.5mm]

$v(3 \rightarrow 0)$   & 0.0206307  & 0.0133490  & 0.0116159  & 0.0106566  & 0.0099446 \\
$v(3 \rightarrow 1)$   & 0.4750377  & 0.3122354  & 0.2759806  & 0.2571594  & 0.2437222 \\
$v(3 \rightarrow 2)$   & 5.1223032  & 3.4625057  & 3.1462448  & 3.0126889  & 2.9330046 \\
$v(3 \rightarrow 3)$   & 0.0003849  & 0.0036817  & 0.0132315  & 0.0322462  & 0.0637063 \\
$v(3 \rightarrow 4)$   & 1.9695305  & 2.2963623  & 2.3790954  & 2.3740069  & 2.3245608 \\
$v(3 \rightarrow 5)$   & 0.4237617  & 0.5080533  & 0.5414953  & 0.5561622  & 0.5608463 \\[1.5mm]

$v(4 \rightarrow 0)$   & 0.0031046  & 0.0019894  & 0.0017143  & 0.0015575  & 0.0014393 \\
$v(4 \rightarrow 1)$   & 0.0720442  & 0.0466771  & 0.0406696  & 0.0373586  & 0.0349071 \\
$v(4 \rightarrow 2)$   & 0.8467202  & 0.5572407  & 0.4931396  & 0.4600513  & 0.4365099 \\
$v(4 \rightarrow 3)$   & 6.1717091  & 4.1766144  & 3.7990409  & 3.6411496  & 3.5477756 \\
$v(4 \rightarrow 4)$   & 0.0003739  & 0.0035759  & 0.0128487  & 0.0313054  & 0.0618278 \\
$v(4 \rightarrow 5)$   & 2.2148342  & 2.5784701  & 2.6669731  & 2.6565183  & 2.5961584 \\[1.5mm]

$v(5 \rightarrow 0)$   & 0.0005732  & 0.0003647  & 0.0003119  & 0.0002814  & 0.0002581 \\
$v(5 \rightarrow 1)$   & 0.0132714  & 0.0085157  & 0.0073482  & 0.0066849  & 0.0061863 \\
$v(5 \rightarrow 2)$   & 0.1574123  & 0.1021202  & 0.0890924  & 0.0819442  & 0.0766644 \\
$v(5 \rightarrow 3)$   & 1.2574929  & 0.8286297  & 0.7342083  & 0.6857528  & 0.6513999 \\
$v(5 \rightarrow 4)$   & 6.9587865  & 4.7146485  & 4.2928548  & 4.1182241  & 4.0158736 \\
$v(5 \rightarrow 5)$   & 0.0003622  & 0.0034636  & 0.0124431  & 0.0303096  & 0.0598417 \\
\hline\hline
\end{tabular}
\end{center}
\end{table}

\begin{table}[h]
\caption{Mean values of various operators for the rovibrational states $(L,v)$ in the T$_2^+$ molecular ion (in a.u.).} \label{T2plus-av}
\begin{center}
\begin{tabular}{c@{\hspace{3mm}}c@{\hspace{3mm}}c@{\hspace{3mm}}c@{\hspace{3mm}}c@{\hspace{3mm}}c@{\hspace{3mm}}c@{\hspace{3mm}}c}
\hline\hline
\vrule width0pt height 11pt
 & $v$  & $\langle\delta(\textbf{r}_1)\rangle$  & $\langle\textbf{p}^4_e\rangle$ & $\langle\textbf{P}^4_1\rangle$ & $R_{te}$ & $R_{tt}$ & $Q_{te}$ \\
\hline
\vrule width0pt height 11pt
$L$ = 0 & 0		&0.20815132766		&6.3301383228		&225.81768163		&1.177858188		&8.156876566		&$-$0.1353386759     	\\
		& 1		&0.20490736213		&6.2335933409		&1025.5963784		&1.160243776		&23.53095592		&$-$0.1334583237  		\\
		& 2		&0.20180643403		&6.1415292480		&2473.9191816		&1.143525406		&37.51331826		&$-$0.1316634315  		\\
		& 3		&0.19884301510		&6.0537788679		&4453.1974638		&1.127673764		&50.17741994		&$-$0.1299511658  		\\
		& 4		&0.19601197899		&5.9701910830		&6856.3167364		&1.112662174		&61.59028274		&$-$0.1283189134  		\\
		& 5		&0.19330825362		&5.8906028212		&9585.5390045		&1.098466270		&71.81294372		&$-$0.1267632470  		\\[1.5mm]
$L$ = 1	&0		&0.20806806060		&6.3275481538		&234.65757447		&1.177416459		&8.397221570		&$-$0.1352845191  		\\
		&1		&0.20482719779		&6.2311019604		&1048.9201758		&1.159820032		&23.75127087		&$-$0.1334061259  		\\
		&2		&0.20172925209 		&6.1391314897		&2509.5998826		&1.143119101		&37.71465023		&$-$0.1316130670  		\\
		&3		&0.19876880131		&6.0514767283		&4499.2828413		&1.127284425		&50.36072666		&$-$0.1299028710  		\\
		&4		&0.19594047900		&5.9679683868		&6911.0087949		&1.112289269		&61.75643149		&$-$0.1282720867  		\\
		&5		&0.19323988121		&5.8884897392		&9647.1921907		&1.098109826		&71.96275502		&$-$0.1267193881  		\\[1.5mm]
$L$ = 2	&0		&0.20790192543		&6.3223804450		&253.76476237		&1.176535079		&8.875956163		&$-$0.1351764724 		\\
		&1		&0.20466725408		&6.2261313424		&1096.8673403		&1.158974555		&24.19002939		&$-$0.1333019827 		\\
		&2		&0.20134537152		&6.1343547384		&2582.1433767		&1.142308481		&38.11552435		&$-$0.1315129307 		\\
		&3		&0.19839935702		&6.0468821601		&4592.5269388		&1.126507623		&50.72562820		&$-$0.1298060980 		\\
		&4		&0.19579808861		&5.9635610113		&7021.3739214		&1.111545512		&62.08711445		&$-$0.1281794270 		\\
		&5		&0.19310276420		&5.8842430571		&9771.3734862		&1.097398074		&72.26079786		&$-$0.1266296115 		\\[1.5mm]
$L$ = 3 &0		&0.20765372596		&6.3146609431		&285.95503192		&1.175218178		&9.589201815		&$-$0.1350151123		\\
		&1		&0.20442832789		&6.2187068901		&1171.9991134		&1.157711356		&24.84351987		&$-$0.1331465149 		\\
		&2		&0.20134537152		&6.1272152705		&2693.8752375		&1.141097365		&38.71238622 		&$-$0.1313630585 		\\
		&3		&0.19839935702		&6.0400194407		&4735.0367475		&1.125347164		&51.26871854		&$-$0.1296618856 		\\
		&4		&0.19558521151		&5.9569663686		&7189.3088412		&1.110434348		&62.57904655		&$-$0.1280404605 		\\
		&5		&0.19289814884		&5.8779128055		&9959.7950893		&1.096335021		&72.70393644 		&$-$0.1264961682 		\\[1.5mm]
$L$ = 4	&0		&0.20732464333		&6.3044274632		&335.35597472		&1.173471887		&10.53122050		&$-$0.1348012493 		\\
		&1		&0.20411154982		&6.2088649942		&1278.0644578		&1.156036370		&25.70625322		&$-$0.1329404519 		\\
		&2		&0.20104051788		&6.1177528903		&2848.1949892		&1.139491567		&39.49998419		&$-$0.1311646356		\\
		&3		&0.19810604393		&6.0309247429		&4929.8896089		&1.123808596		&51.98497524		&$-$0.1294708469 		\\
		&4		&0.19530309749		&5.9482292316		&7417.5916452		&1.108961269		&63.22742191		&$-$0.1278566161 		\\
		&5		&0.19262692747		&5.8695237128		&10214.960197		&1.094925830		&73.28757857		&$-$0.1263193748 		\\[1.5mm]
$L$ = 5	&0		&0.20691623601		&6.2917297093		&407.29565446		&1.171304261		&11.69450764		&$-$0.1345359738 		\\
		&1		&0.20371843571		&6.1966536647		&1419.8923833		&1.153957402		&26.77105351		&$-$0.1326848715 		\\
		&2		&0.20066220613		&6.1060128421		&3049.4710411		&1.137498624		&40.47145500		&$-$0.1309184836 		\\
		&3		&0.19774210049		&6.0196424218		&5181.0276540		&1.121899288		&52.86783165		&$-$0.1292339503 		\\
		&4		&0.19495306988		&5.9373913615		&7709.7712139		&1.107133381		&64.02596164		&$-$0.1276286468		\\
		&5		&0.19229044515		&5.8591185343		&10540.049995		&1.093177405		&74.00571285		&$-$0.1261001867 		\\
\hline\hline
\end{tabular}
\end{center}
\end{table}

\begin{table}[t]
\caption{Mean values of various operators for the rovibrational states $(L,v)$ in the DT$^+$ molecular ion (in a.u.).} \label{DTplus-av}
\begin{center}
\begin{tabular}{c@{\hspace{3mm}}c@{\hspace{3mm}}c@{\hspace{3mm}}c@{\hspace{3mm}}c@{\hspace{3mm}}c@{\hspace{3mm}}c}
\hline\hline
\vrule width0pt height 11pt
  & $v$  & $\langle\delta(\textbf{r}_d)\rangle$     &$\langle\delta(\textbf{r}_p)\rangle$     & $\langle\textbf{p}^4_e\rangle$      & $\langle\textbf{P}^4_d\rangle$    & $\langle\textbf{P}^4_t\rangle$     \\
\hline
\vrule width0pt height 11pt
$L = 0$ &0		&0.20798062505		&0.20787917082		&6.3232479907		&182.35056287		&182.39001528				\\
		&1		&0.20437585095		&0.20427282570		&6.2159760442		&818.08651728		&818.25119116				\\
		&2		&0.20094859620		&0.20084367246		&6.1142608870		&1955.5534029		&1955.9336681				\\
		&3		&0.19769127061		&0.19758407965		&6.0178721554		&3491.8394348		&3492.5070856				\\
		&4		&0.19459693666		&0.19448704675		&5.9266006043		&5334.1729004		&5335.1829486				\\
		&5		&0.19165911234		&0.19154603205		&5.8402496573		&7398.7412779		&7400.1335775				\\[1.5mm]
$L = 1$	&0		&0.20787690907		&0.20777545518		&6.3200219755		&190.29843934		&190.33965948				\\
		&1		&0.20427645295		&0.20417342315		&6.2128873506		&838.74873617		&838.91747070				\\
		&2		&0.20085334472		&0.20074841903		&6.1113040388		&1986.8363471		&1987.2225943				\\
		&3		&0.19760006366		&0.19749286411		&6.0150445270		&3531.8396030		&3532.5148000				\\
		&4		&0.19450956462		&0.19439969505		&5.9238950769		&5381.1552609		&5382.1741125				\\
		&5		&0.19157559207		&0.19146252250		&5.8376684527		&7451.1207645		&7452.5228573				\\[1.5mm]
$L = 2$	&0		&0.20767010031		&0.20756864770		&6.3135898446		&207.61051177		&207.65552189				\\
		&1		&0.20407825941		&0.20397521843		&6.2067291215		&881.34811539		&881.52520038				\\
		&2		&0.20066348048		&0.20055852268		&6.1054110295		&2050.5477834		&2050.9461870				\\
		&3		&0.19741820484		&0.19731096428		&6.0094062054		&361.28667505		&3613.5572240				\\
		&4		&0.19433551130		&0.19422557251		&5.9185059565		&5476.0379777		&5477.0745996				\\
		&5		&0.19140903065		&0.19129590082		&5.8325183309		&7556.6962851		&7558.1180737				\\[1.5mm]
$L = 3$	&0		&0.20736144102		&0.20725999016		&6.3039913318		&237.07095898		&237.12228166				\\
		&1		&0.20378247008		&0.20367942065		&6.1975399415		&948.38779763		&948.57797209				\\
		&2		&0.20038010360		&0.20027512803		&6.0966170145		&2148.9336326		&2149.3507762				\\
		&3		&0.19714682793		&0.19703954539		&6.0009944019		&3736.9307561		&3737.6445955				\\
		&4		&0.19407574805		&0.19396575024		&5.9104646427		&5620.6133472		&5621.6770286				\\
		&5		&0.19116053362		&0.19104732446		&5.8248368966		&7717.0602601		&7718.5119483				\\[1.5mm]
$L = 4$	&0		&0.20695276646		&0.20685131658		&6.2912850632		&282.73714458		&282.79803146				\\
		&1		&0.20339086268		&0.20328779667		&6.1853763992		&1043.5078994		&1043.7165563				\\
		&2		&0.20000495952		&0.19989995011		&6.0849775315		&2285.2522430		&2285.6952955				\\
		&3		&0.19678759227		&0.19668026093		&5.9898619427		&3906.9399513		&3907.6857696		 		\\
		&4		&0.19373191582		&0.19362183955		&5.8998234256		&5817.4679498		&5818.5684431				\\
		&5		&0.19083164201		&0.19071832604		&5.8146728757		&7934.5023469		&7935.9945499				\\[1.5mm]
$L = 5$	&0		&0.20644647749		&0.20634502821		&6.2755477676		&349.80306411		&349.87769999				\\
		&1		&0.20290575685		&0.20280267138		&6.1703127066		&1171.3541215		&1171.5874879				\\
		&2		&0.19954028822		&0.19943523563		&6.0705642100		&2463.6479109		&2464.1247846				\\
		&3		&0.19634267058		&0.19623527364		&5.9760779905		&4126.5789886		&4127.3660582		 		\\
		&4		&0.19330611635		&0.19319593979		&5.8866493653		&6069.8640651		&6071.0117071				\\
		&5		&0.19042439216		&0.19031094111		&5.8020913123		&8211.8955253		&8213.4393726				\\
\hline\hline
\end{tabular}
\end{center}
\end{table}

\begin{table}[t]\addtocounter{table}{-1}
\caption{Mean values of various operators for the rovibrational states $(L,v)$ in the DT$^+$ molecular ion (in a.u.). {\it(Continued)}} \label{DTplus-av_c}
\begin{center}
\begin{tabular}{c@{\hspace{3mm}}c@{\hspace{3mm}}c@{\hspace{3mm}}c@{\hspace{3mm}}c@{\hspace{3mm}}c@{\hspace{3mm}}c}
\hline\hline
\vrule width0pt height 11pt
        & $v$    & $R_{dt}$ 		 & $R_{te}$  	& $R_{de}$    & $Q_{de}$  & $Q_{te}$	  \\
\hline
\vrule width0pt height 11pt
$L = 0$	& 0			&7.271242711		&1.177372661		&1.175921538		&$-$0.1352437844 		&$-$0.1351536481 		\\
		& 1			&20.87895645		&1.158588274		&1.155581519		&$-$0.1331546728 		&$-$0.1330638073  		\\
		& 2			&33.10841449		&1.140846803		&1.136422873		&$-$0.1311717620 		&$-$0.1310799167  		\\
		& 3			&44.03997401		&1.124111877		&1.118401897		&$-$0.1292910793 		&$-$0.1291979817 		\\
		& 4			&53.74613235		&1.108350884		&1.101479317		&$-$0.1275091354 		&$-$0.1274144288 		\\
		& 5			&62.29212023		&1.093534730		&1.085619971		&$-$0.1258223222 		&$-$0.1257256621		\\[1.5mm]
$L = 1$	& 0			&7.510315827		&1.176846371		&1.175347634		&$-$0.1351763223 		&$-$0.1350861962 		\\
		& 1			&21.09579419		&1.158085563		&1.155034035		&$-$0.1330899519 		&$-$0.1329990899  		\\
		& 2			&33.30430437		&1.140366902		&1.135900919		&$-$0.1311096331 		&$-$0.1310178154  		\\
		& 3			&44.21608041		&1.123654115		&1.117904682		&$-$0.1292316053 		&$-$0.1291385290 		\\
		& 4			&53.90350480		&1.107914604		&1.101006099		&$-$0.1274519621 		&$-$0.1273574051 		\\
		& 5			&62.43170287		&1.093119406		&1.085170200		&$-$0.1257678916 		&$-$0.1256713146		\\[1.5mm]
$L = 2$	& 0			&7.986028949		&1.175796817		&1.174203261		&$-$0.1350418264 		&$-$0.1349517226 		\\
		& 1			&21.52715287		&1.157083065		&1.153942382		&$-$0.1329609231 		&$-$0.1328700623  		\\
		& 2			&33.69387755		&1.139409944		&1.134860225		&$-$0.1309859813 		&$-$0.1308941111  		\\
		& 3			&44.56619112		&1.122741309		&1.116913336		&$-$0.1291130219 		&$-$0.1290198957 		\\
		& 4			&54.21624805		&1.107044739		&1.100062695		&$-$0.1273385161 		&$-$0.1272438402 		\\
		& 5			&62.70895892		&1.092291374		&1.084273419		&$-$0.1256591674 		&$-$0.1255625691		\\[1.5mm]
$L = 3$	& 0			&8.693566042		&1.174230010		&1.172495234		&$-$0.1348411676 		&$-$0.1347510982 		\\
		& 1			&22.16844786		&1.155586583		&1.152313144		&$-$0.1327684259 		&$-$0.1326775987  		\\
		& 2			&34.27276842		&1.137981517		&1.133307138		&$-$0.1308013942 		&$-$0.1307095675  		\\
		& 3			&45.08614797		&1.121378885		&1.115434022		&$-$0.1289361507 		&$-$0.1288430210 		\\
		& 4			&54.68039950		&1.105746487		&1.098655011		&$-$0.1271691477 		&$-$0.1270744576 		\\
		& 5			&63.12011420		&1.091055662		&1.082935470		&$-$0.1254971270 		&$-$0.1254005002		\\[1.5mm]
$L = 4$	& 0			&9.625826574		&1.172154848		&1.170233643		&$-$0.1345756236 		&$-$0.1344855971 		\\
		& 1			&23.01292273		&1.153604710		&1.150156057		&$-$0.1325137096 		&$-$0.1324229084  		\\
		& 2			&35.03454605		&1.136089929		&1.131251066		&$-$0.1305571855 		&$-$0.1304653747  		\\
		& 3			&45.76982872		&1.119574872		&1.113475813		&$-$0.1287021553 		&$-$0.1286090447 		\\
		& 4			&55.29012996		&1.104027617		&1.096791839		&$-$0.1269451055 		&$-$0.1268503994 		\\
		& 5			&63.65961719		&1.089419782		&1.081164822		&$-$0.1252827998 		&$-$0.1251861341		\\[1.5mm]
$L = 5$	& 0			&10.77356847		&1.169582982		&1.167431692		&$-$0.1342468639 		&$-$0.1341568918 		\\
		& 1			&24.05178854		&1.151148695		&1.147483867		&$-$0.1321983863 		&$-$0.1321076263  		\\
		& 2			&35.97085002		&1.133746046		&1.128704314		&$-$0.1302549005 		&$-$0.1301631120  		\\
		& 3			&46.60927850		&1.117339764		&1.111050602		&$-$0.1284125550 		&$-$0.1283194602 		\\
		& 4			&56.03786914		&1.101898281		&1.094484666		&$-$0.1266678656 		&$-$0.1265731355 		\\
		& 5			&64.32026245		&1.087393556		&1.078972585		&$-$0.1250176248 		&$-$0.1249209091		\\
\hline\hline
\end{tabular}
\end{center}
\end{table}

\clearpage

\section{Hyperfine structure}

\begin{table}[h]
\caption{Coefficients $E_i$ of the effective spin Hamiltonian (10) for the DT$^+$ molecular ion (in MHz), $a[b]=a\times10^b$.}
\begin{tabular}{c@{\hspace{2mm}}c@{\hspace{2mm}}c@{\hspace{2mm}}c@{\hspace{2mm}}c@{\hspace{2mm}}c@{\hspace{2mm}}c@{\hspace{2mm}}c@{\hspace{2mm}}c@{\hspace{2mm}}c@{\hspace{2mm}}c}
\hline\hline
  $L$  & $v$ & $E_1$ & $E_2$ & $E_3$ & $E_4$ & $E_5$ & $E_6$ & $E_7$ & $E_8$ & $E_9$ \\
\hline
0 & 0 & --- & --- & --- & 991.60054 & 142.63716 & --- & --- & --- & --- \\
0 & 1 & --- & --- & --- & 974.41386 & 140.16265 & --- & --- & --- & --- \\
0 & 2 & --- & --- & --- & 958.07355 & 137.80972 & --- & --- & --- & --- \\
0 & 3 & --- & --- & --- & 942.54342 & 135.57314 & --- & --- & --- & --- \\
0 & 4 & --- & --- & --- & 927.79040 & 133.44809 & --- & --- & --- & --- \\
0 & 5 & --- & --- & --- & 913.78358 & 131.43010 & --- & --- & --- & --- \\[1.5mm]
1 & 0 & 17.93543 & $-$0.18716[$-$1] & $-$0.53588[$-$2] &  991.10605 &  142.56599 &  9.26118 &  1.33286 & $-$0.32813[$-$2] &  0.56718[$-$2] \\
1 & 1 & 17.21975 & $-$0.18369[$-$1] & $-$0.52076[$-$2] &  973.93996 &  140.09444 &  8.87897 &  1.27785 & $-$0.31923[$-$2] &  0.56722[$-$2] \\
1 & 2 & 16.52644 & $-$0.18002[$-$1] & $-$0.50562[$-$2] &  957.61942 &  137.74436 &  8.50937 &  1.22466 & $-$0.31027[$-$2] &  0.56552[$-$2] \\
1 & 3 & 15.85367 & $-$0.17615[$-$1] & $-$0.49043[$-$2] &  942.10857 &  135.51055 &  8.15138 &  1.17314 & $-$0.30126[$-$2] &  0.56219[$-$2] \\
1 & 4 & 15.19971 & $-$0.17209[$-$1] & $-$0.47519[$-$2] &  927.37383 &  133.38816 &  7.80404 &  1.12316 & $-$0.29220[$-$2] &  0.55730[$-$2] \\
1 & 5 & 14.56273 & $-$0.16785[$-$1] & $-$0.45990[$-$2] &  913.38537 &  131.37280 &  7.46645 &  1.07457 & $-$0.28307[$-$2] &  0.55093[$-$2] \\[1.5mm]
2 & 0 & 17.89054 & $-$0.18644[$-$1] & $-$0.53421[$-$2] &  990.12003 &  142.42409 &  2.20078 &  0.31673 & $-$0.77866[$-$3] &  0.13423[$-$2] \\
2 & 1 & 17.17639 & $-$0.18298[$-$1] & $-$0.51912[$-$2] &  972.99502 &  139.95844 &  2.10993 &  0.30366 & $-$0.75749[$-$3] &  0.13423[$-$2] \\
2 & 2 & 16.48456 & $-$0.17931[$-$1] & $-$0.50399[$-$2] &  956.71419 &  137.61406 &  2.02207 &  0.29102 & $-$0.73620[$-$3] &  0.13383[$-$2] \\
2 & 3 & 15.81309 & $-$0.17545[$-$1] & $-$0.48884[$-$2] &  941.24151 &  135.38574 &  1.93697 &  0.27877 & $-$0.71479[$-$3] &  0.13304[$-$2] \\
2 & 4 & 15.16046 & $-$0.17139[$-$1] & $-$0.47362[$-$2] &  926.54399 &  133.26868 &  1.85440 &  0.26689 & $-$0.69324[$-$3] &  0.13188[$-$2] \\
2 & 5 & 14.52472 & $-$0.16716[$-$1] & $-$0.45835[$-$2] &  912.59125 &  131.25848 &  1.77414 &  0.25533 & $-$0.67154[$-$3] &  0.13037[$-$2] \\[1.5mm]
3 & 0 & 17.82356 & $-$0.18538[$-$1] & $-$0.53172[$-$2] &  988.64842 &  142.21230 &  1.02406 &  0.14738 & $-$0.36156[$-$3] &  0.62072[$-$3] \\
3 & 1 & 17.11169 & $-$0.18192[$-$1] & $-$0.51666[$-$2] &  971.58477 &  139.75548 &  0.98177 &  0.14130 & $-$0.35170[$-$3] &  0.62075[$-$3] \\
3 & 2 & 16.42203 & $-$0.17825[$-$1] & $-$0.50157[$-$2] &  955.36312 &  137.41961 &  0.94087 &  0.13541 & $-$0.34180[$-$3] &  0.61888[$-$3] \\
3 & 3 & 15.75272 & $-$0.17440[$-$1] & $-$0.48645[$-$2] &  939.94766 &  135.19950 &  0.90125 &  0.12971 & $-$0.33183[$-$3] &  0.61521[$-$3] \\
3 & 4 & 15.10197 & $-$0.17035[$-$1] & $-$0.47128[$-$2] &  925.30550 &  133.09040 &  0.86280 &  0.12417 & $-$0.32180[$-$3] &  0.60984[$-$3] \\
3 & 5 & 14.46809 & $-$0.16613[$-$1] & $-$0.45604[$-$2] &  911.40647 &  131.08791 &  0.82543 &  0.11880 & $-$0.31170[$-$3] &  0.60284[$-$3] \\[1.5mm]
4 & 0 & 17.73492 & $-$0.18397[$-$1] & $-$0.52844[$-$2] &  986.69996 &  141.93189 &  0.59617 &  0.08580 & $-$0.20990[$-$3] &  0.35839[$-$3] \\
4 & 1 & 17.02607 & $-$0.18052[$-$1] & $-$0.51342[$-$2] &  969.71768 &  139.48676 &  0.57154 &  0.08226 & $-$0.20416[$-$3] &  0.35841[$-$3] \\
4 & 2 & 16.33927 & $-$0.17686[$-$1] & $-$0.49838[$-$2] &  953.57452 &  137.16218 &  0.54772 &  0.07883 & $-$0.19839[$-$3] &  0.35732[$-$3] \\
4 & 3 & 15.67269 & $-$0.17301[$-$1] & $-$0.48331[$-$2] &  938.23491 &  134.95298 &  0.52463 &  0.07551 & $-$0.19259[$-$3] &  0.35520[$-$3] \\
4 & 4 & 15.02454 & $-$0.16898[$-$1] & $-$0.46818[$-$2] &  923.66619 &  132.85443 &  0.50223 &  0.07228 & $-$0.18675[$-$3] &  0.35208[$-$3] \\
4 & 5 & 14.39315 & $-$0.16477[$-$1] & $-$0.45300[$-$2] &  909.83840 &  130.86217 &  0.48046 &  0.06915 & $-$0.18086[$-$3] &  0.34802[$-$3] \\[1.5mm]
5 & 0 & 17.62515 & $-$0.18224[$-$1] & $-$0.52438[$-$2] &  984.28610 &  141.58450 &  0.39047 &  0.05620 & $-$0.13700[$-$3] &  0.23233[$-$3] \\
5 & 1 & 16.92005 & $-$0.17879[$-$1] & $-$0.50942[$-$2] &  967.40481 &  139.15389 &  0.37433 &  0.05387 & $-$0.13324[$-$3] &  0.23234[$-$3] \\
5 & 2 & 16.23681 & $-$0.17514[$-$1] & $-$0.49444[$-$2] &  951.35909 &  136.84331 &  0.35871 &  0.05163 & $-$0.12946[$-$3] &  0.23163[$-$3] \\
5 & 3 & 15.57361 & $-$0.17131[$-$1] & $-$0.47942[$-$2] &  936.11363 &  134.64765 &  0.34358 &  0.04945 & $-$0.12565[$-$3] &  0.23024[$-$3] \\
5 & 4 & 14.92867 & $-$0.16729[$-$1] & $-$0.46436[$-$2] &  921.63609 &  132.56220 &  0.32890 &  0.04733 & $-$0.12183[$-$3] &  0.22821[$-$3] \\
5 & 5 & 14.30035 & $-$0.16309[$-$1] & $-$0.44924[$-$2] &  907.89673 &  130.58264 &  0.31462 &  0.04528 & $-$0.11798[$-$3] &  0.22557[$-$3] \\
\hline\hline
\end{tabular}
\end{table}

\begin{table}[t]
\begin{center}
\caption{Coefficients of the effective spin Hamiltonian (12) for the T$_2^+$ molecular ion (in MHz). $a[b]=a\times10^{b}$.} \label{heff-tab}
\begin{tabular}{cc@{\hspace{2mm}}@{\hspace{3mm}}c@{\hspace{3mm}}c@{\hspace{2.5mm}}c@{\hspace{2.5mm}}c@{\hspace{2.5mm}}c}
\hline\hline \vrule width0pt height10pt
$L$ & $v$ & $b_F$ & $c_e$ & $c_I$ & $d_1$ & $d_2$ \\
\hline \vrule width0pt height11pt
1 & 0 & 992.017 & 14.3979 & $-$1.500[$-$02] & 139.266 & $-$0.3426 \\
1 & 1 & 976.566 & 13.8831 & $-$1.476[$-$02] & 134.116 & $-$0.3343 \\
1 & 2 & 961.795 & 13.3829 & $-$1.449[$-$02] & 129.119 & $-$0.3260 \\
1 & 3 & 947.681 & 12.8961 & $-$1.422[$-$02] & 124.265 & $-$0.3176 \\
1 & 4 & 934.196 & 12.4218 & $-$1.393[$-$02] & 119.543 & $-$0.3092 \\
1 & 5 & 921.320 & 11.9590 & $-$1.364[$-$02] & 114.943 & $-$0.3007 \\
\vrule width0pt height10.5pt
3 & 0 & 990.042 & 14.3259 & $-$1.489[$-$02] & 138.728 & $-$0.3403 \\
3 & 1 & 974.664 & 13.8133 & $-$1.464[$-$02] & 133.595 & $-$0.3321 \\
3 & 2 & 959.965 & 13.3152 & $-$1.438[$-$02] & 128.615 & $-$0.3237 \\
3 & 3 & 945.919 & 12.8305 & $-$1.411[$-$02] & 123.776 & $-$0.3154 \\
3 & 4 & 932.502 & 12.3581 & $-$1.382[$-$02] & 119.068 & $-$0.3070 \\
3 & 5 & 919.691 & 11.8971 & $-$1.353[$-$02] & 114.482 & $-$0.2985 \\
\vrule width0pt height10.5pt
5 & 0 & 986.526 & 14.1979 & $-$1.469[$-$02] & 137.769 & $-$0.3363 \\
5 & 1 & 971.279 & 13.6892 & $-$1.444[$-$02] & 132.667 & $-$0.3281 \\
5 & 2 & 956.708 & 13.1948 & $-$1.418[$-$02] & 127.715 & $-$0.3198 \\
5 & 3 & 942.786 & 12.7137 & $-$1.391[$-$02] & 122.904 & $-$0.3115 \\
5 & 4 & 929.488 & 12.2447 & $-$1.362[$-$02] & 118.222 & $-$0.3031 \\
5 & 5 & 916.794 & 11.7870 & $-$1.333[$-$02] & 113.661 & $-$0.2947 \\
\hline
\end{tabular}\\[-1pt]
\begin{tabular}{@{\hspace{4pt}}c@{\hspace{1mm}}c@{\hspace{6mm}}c@{\hspace{8mm}}c@{\hspace{1mm}}c@{\hspace{6mm}}c@{\hspace{1mm}}}
\vrule width0pt height10pt
$L$ & $v$ & $c_e$ & $L$ & $v$ & $c_e$ \\
\hline \vrule width0pt height10.5pt
2 & 0 & 14.3690 & 4 & 0 & 14.2687 \\
2 & 1 & 13.8551 & 4 & 1 & 13.7579 \\
2 & 2 & 13.3558 & 4 & 2 & 13.2615 \\
2 & 3 & 12.8698 & 4 & 3 & 12.7783 \\
2 & 4 & 12.3963 & 4 & 4 & 12.3075 \\
2 & 5 & 11.9341 & 4 & 5 & 11.8479 \\
\hline\hline
\end{tabular}
\vspace{-2mm}
\end{center}
\end{table}

\end{document}